\documentclass[11pt,dvips]{article}
cm \voffset = -2truecm

\usepackage{graphicx}
\usepackage{amssymb}
\usepackage{latexsym}
\begin{document}
\begin{titlepage}
\setcounter{page}{1}
\renewcommand{\thefootnote}{\fnsymbol{footnote}}

\vspace{5mm}
\begin{center}
 {\Large \bf
Generalized fermion algebra}

\vspace{0.5cm}
{\bf Won Sang Chung$^1$}{\footnote { email: {\sf
mimip4444@hanmail.net}}} and {\bf Mohammed Daoud$^{2,3}$}{\footnote {
email:
{\sf m$_-$daoud@hotmail.com}}}\\
\vspace{0.5cm} {\em $^1$ Department of Physics and Research
Institute of Natural Science,\\ College of Natural Science,
Gyeongsang National
University, Jinju 660-701, Korea}\\

\vspace{0.5cm} {\em $^2$ Facult\'e des Sciences,
D\'epartement de Physique, \\Université Ibnou Zohr, Agadir, Morocco}\\

\vspace{0.5cm} {\em $^3$ Abdus Salam International Centre for Theoretical Physics, \\Strada Costiera, 11 I-34151 Trieste, Italy}\\

\vspace{4cm}
\begin{abstract}
A one-parameter generalized fermion algebra ${\cal B}_{\kappa}(1)$ is introduced. The
Fock representation is studied. The associated coherent states are constructed and the polynomial representation, in the Bargmann sense, is derived.
A special attention is devoted to the limiting case $\kappa \rightarrow 0$ where the fermionic coherent states,
labeled by Grassmann variables, are obtained. The physical relevance
of the algebra is illustrated throughout  Calogero-Sutherland system.

\end{abstract}
\end{center}
\end{titlepage}
\newpage
\section{Introduction and motivations}

Over the last two decades, major effort has been directed towards
the generalized Weyl-Heisenberg algebras. In connection with
the theory of quantum algebras and in the spirit of the pioneering works \cite{Arik,Mac,Bied} a wide class of non linear bosonic algebras
were investigated from many perspectives and for different purposes. The common feature of all these
generalizations  is their structural similarity and subsequently they can be described in
a unified mathematical framework \cite{Dask}. Indeed, any generalized Weyl-Heisenberg algebra can be defined by the
basic structure relations
$$a^+ a^- = F_-(N) \quad a^- a^+ = F_-(N+1) \quad [N , a^{\pm}] =  \pm a^{\pm} $$
where the structure
function $F_-(N)$ ($N$ stands for the usual number operator) defined as the product of the
creation and annihilation operators  $a^+$  and  $a^-$. Equivalently, the structure relations of the  extended
Weyl-Heisenberg write
$$ [a^+ ,  a^- ] = F_-(N+1) - F_-(N) \equiv  G_-(N) \qquad [N , a^{\pm}] =  \pm a^{\pm}. $$
The function $F(N)$  characterizes the generalization or deformation
scheme. Several possible generalizations can be defined, each one might be useful for different purposes (see for instance \cite{Fu,Quesne1,Quesne2,Antoine,Gazeau}).
For instance, the polynomial  Weyl-Heisenberg
algebras,  corresponding to the case where the structure function  $F_-(N)$  is
a polynomial in $N$,   are of special interest for quantum
systems with non-linear discrete spectrum \cite{Daoud2,Daoud3,Daoud4,Daoud5}.
In this context, an interesting  situation corresponds to the case where $G_-(N) $ is linear in $N$ :
  $$G_-(N) =   1  + 2\kappa N $$
 The usual harmonic oscillator is recovered when $\kappa = 0$.
This algebra includes $su(2)$ for $\kappa < 0$ , $su(1,1)$ for $\kappa > 0$ and it related to some exactly solvable potentials like Morse and P\"oschl-Teller ones \cite{Daoud3}.

In other
hand, paralleling the extension or deformation
of  standard structure relations defining  the bosonic oscillator algebra, several extensions
schemes of the fermionic algebra were studied in \cite{Bona}. In particular, the authors concluded that
all deformed fermionic algebras, in the context of the $q$-deformations inherited from quantum Lie algebras,  are isomorphically equivalent to the non-deformed fermionic algebra
when the deformation parameter $q$ is generic. For $q$ a root of unity, deformation schemes of fermions provide a nice mathematical framework
to define objects interpolating between bosons and fermions (see \cite{Daoud1} and references therein).  In this respect, motivated by the polynomial extension of
the usual bosonic algebra presented briefly above, we shall mainly focus on the  generalization of the usual
fermion algebra given by the following basic structure relations
$$f^+ f^- = F_+(N) \quad f^- f^+ = F_+(N+1) \quad [N , f^{\pm}] =  \pm f^{\pm} $$
which can be re-equated  also as
$$\{f^+ ,  f^-\} = F_+(N+1) + F_+(N) \equiv  G_+(N) \qquad [N , f^{\pm}] =  \pm f^{\pm} .$$

In Section 2, we consider a particular, but nevertheless very large
class of extended fermion algebra.  We shall pay attention to the one parameter algebra where the function $G_+(N)$ is
linear in the number operator $N$. The Fock representation space is explicitly defined.  Section 3  deals
with the connection with this particular class of generalized fermion algebra and the
two-body Calogero-Sutherland model. We also discuss the bosonization procedure, leading to  the generalized
algebra ${\cal A}_{\kappa}(1)$ introduced in \cite{Daoud3}, by using a suitable  boson mapping in terms of generalized  fermion generators. In Section 4,
an analytical representation of the algebra is derived. It is based on the associated coherent states. Finally, normal ordering process is discussed in Section 5. The expressions of the analogue of Striling and Bell operators are obtained.
Concluding remarks and possible prolongations of the present work close this paper.

\section{Generalized fermion algebra }

As mentioned  above,  this work focuses on the special class of  the generalized or extended fermion algebra where $G_+(N)$
is linear in the number operator. Analogously to the generalized boson or Weyl-Heisenberg algebra ${\cal A}_{\kappa}(1)$ discussed
in \cite{Daoud3},  it will be denoted by ${\cal B}_{\kappa}(1)$.  In this case, the generators $\{f^+
, f^- ,N \}$, satisfy  the structure
relations
\begin{eqnarray}
\{f^- , f^+\} = \mathbb{I} + 2\kappa N  \qquad [N, f^-] = -f^-,\qquad [N, f^+ ] = + f^+,
\end{eqnarray}
where $\mathbb{I}$ is the unity operator and $\kappa$ is a real parameter characterizing the deviation from
the usual fermion algebra which is recovered for   $\kappa = 0$. It is interesting to note that the generalized fermionic operators obey the following
identities
\begin{equation}
f^- ( f^+)^n = (-1)^n ( f^+)^n f^- +  \bigg\{ \frac{ 1- (-1)^n }{2}
( 1 + \kappa + 2 \kappa N ) + (-1)^n \kappa n \bigg\} ( f^+)^{n-1}
\end{equation}
and
\begin{equation}
 ( f^-)^n f^+ = (-1)^n  f^+( f^-)^n  +  ( f^-)^{n-1} \bigg \{ \frac{ 1-
(-1)^n }{2} ( 1 + \kappa + 2 \kappa N ) + (-1)^n \kappa n \bigg\}.
\end{equation}

From these equations, one can see that in the the specific case $\kappa = 0$, the operators $( f^+)^2$ and $(f^-)^2$ belong to the center of the algebra.
They satisfy the nilpotency relation $( f^+)^2 =(f^-)^2 = 0$ that reflects the Pauli exclusion principle. Except this particular situation,
there not exists any integer $n$ such that the ladder operators $f^+$ and $f^-$ are nilpotent. Consequently, for  $\kappa \neq 0$, the representations of generalized
fermion are infinite dimensional.  The Fock representation of this algebra is given by
means of a complete set of orthonormal states $ {\cal F} = \{ \vert n \rangle ,
n \in \mathbb{N}\}$ which are eigenstates of the number operator
$N$, $N\vert n \rangle =n \vert n \rangle$. In this representation,
the vacuum state is defined as $f^- \vert 0 \rangle =0$
 and the ortho-normalized eigenstates are
constructed  by successive applications of the creation operator
$f^+$. We define the actions of creation and annihilation
operators as
\begin{eqnarray}\label{action}
 f^-\vert n \rangle =\sqrt{F_+(n)}   \vert n - 1 \rangle ,\qquad
 f^+\vert n \rangle=\sqrt{F_+(n+1)} \vert n+1  \rangle
\label{action over n}
\end{eqnarray}
where the structure function $F_+(.)$ is an analytic function
satisfying the two following  conditions
\begin{eqnarray}
 F_+(0)=0 \qquad {\rm and} \qquad  F_+(n)>0 , \qquad  n=1,\ldots.
\end{eqnarray}
The action of the operator $G_+(N)$ is given by
$$G_+(N) \vert n \rangle  = ( 1+ 2\kappa n) \vert n \rangle $$

It is simple
to check that, in the Fock space  ${\cal F}$, the  operators $f^+$ and $f^-$ are mutually adjoint, $f^+
= (f^-)^{\dagger}$. It is also easy to verify that the structure function
$F_+(.)$  satisfies the following recursion formula
\begin{eqnarray}
 F_+(n+1) + F_+(n) = G_+(n),
\end{eqnarray}
where  $ G_+(n) = 1+ 2\kappa n $. By simple iteration, we get
\begin{eqnarray}
F_+(n) = (-)^{n-1} \sum_{m=0}^{n-1} (-)^{m} G_+(m),
\end{eqnarray}
and the explicit form
of the structure function $F_+(N)$ is given by

\begin{eqnarray}\label{F_+(N)}
F_+(N) =  \frac{1}{2}(1-(-1)^N) ( 1 + \kappa ( N-1) ) +
\frac{\kappa}{2} N( 1+ (-1)^N).
\end{eqnarray}
It follows that the concrete form of the actions of the creation and annihilation operators (\ref{action}) are
\begin{equation} \label{action-}
f^-|n\rangle = \sqrt{ \frac{1}{2}(1-(-1)^n) ( 1 + \kappa ( n-1) ) +
\frac{\kappa}{2} n ( 1+ (-1)^n ) } | n-1 \rangle
\end{equation}
 \begin{equation} \label{action+}
f^+ |n\rangle
= \sqrt{ \frac{1}{2}(1+(-1)^n) ( 1 + \kappa n ) + \frac{\kappa}{2}(
n+1) ( 1- (-1)^n ) } | n+1 \rangle.
\end{equation}
From Eq.(\ref{F_+(N)}),  the structure
function is given by
$$F_+(n) = \kappa n$$
for $n$ even and for $n$ odd it writes
$$F_+(n) = 1 + \kappa (n-1).$$
It is important to stress that $\kappa$ should be positive to
ensure the positivity of $F_+(n)$. Obviously, the representation space of the algebra ${\cal B}_{\kappa}(1)$ is
infinite dimensional except the limiting case $\kappa = 0$ where the usual fermion algebra
is recovered. In this case, the Fock space ${\cal F}$ is two dimensional
and comprises only the states $\vert 0 \rangle$ and $ \vert 1 \rangle$.


\section{The Hamiltonian and two-body Calogero-Sutherland spectrum }

The algebra ${\cal B}_{\kappa}(1)$ is interesting from two different aspects. The first concerns its relevance
for the study of some one-dimensional exactly solvable
potentials. The second is related to its connection the $\mathbb{Z}_k$ graded Weyl-Heisenberg algebra $W_k$ introduced in \cite{Daoud2}
and the generalized Weyl-Heisenberg algbra ${\cal A}_{\kappa}(1)$ defined in \cite{Daoud3}.

Concerning the first aspect,  the  algebra ${\cal B}_{\kappa}(1)$
can be related to the two-body Calogero-Sutherland model. For this end, we notice that the structure
function $F_+(N) \equiv f^+f^-$ can be  also written as
\begin{equation}\label{F_+(N)1}
F_+(N) = \kappa N + (1 - \kappa) \Pi_1
\end{equation}
where  the operator $\Pi_1$ projects on the odd number Fock states. It is defined by
$$\Pi_1 = \frac{1 - (-)^{N}}{2} $$
which is orthogonal to the operator
$$\Pi_0 = \frac{1 + (-)^{N}}{2} $$
projecting on even Fock  states. This induces a  $\mathbb{Z}_2$ graduation of the Fock space. Accordingly, the products $f^+f^-$ and $f^-f^+$ are
$$f^+f^- = \kappa N + (1 - \kappa) \Pi_1  \qquad f^-f^+ = \kappa (N+1) + (1 - \kappa) \Pi_0 $$
and the commutator between the operators $f^+$ and $f^-$ rewrites
\begin{equation}\label{com-f}
[ f^- ,f^+] = \Pi_0 + (2\kappa - 1) \Pi_1
\end{equation}
which is exactly the $\mathbb{Z}_2$ graded Weyl-Heisenberg algebra $W_2$ \cite{Daoud2}. In view of
this  $\mathbb{Z}_2$ graduation, the algebra ${\cal B}_{\kappa}(1)$  is relevant for the description of the energy
spectra of the one dimensional two-body Calogero-Sutherland model. Indeed, the spectrum of the operator  $f^+f^-$ given by

$$ f^+f^- \vert 2n \rangle = 2\kappa n  ~\vert 2n \rangle \qquad f^+f^- \vert 2n+1 \rangle = (2\kappa n + 1) ~ \vert 2n+1 \rangle  $$
coincides with
 the Hamiltonian corresponding to the Calogero-Sutherland potential (in $x$-representation)
$$
V_0(x,\kappa) = \frac{\kappa^2}{4} x^2 +
                 \frac{1- \kappa^2}{4\kappa^2}
   \frac{1}{x^2} -\bigg(\kappa
 - \frac{1}{2}\bigg).
$$
This agrees with the results of the Ref. \cite{Daoud2}. Furthermore,
the spectrum of the operator $f^-f^+$:
$$ f^-f^+ \vert 2n \rangle = (2\kappa n + 1)~ \vert 2n \rangle \qquad f^-f^+ \vert 2n+1 \rangle = (2\kappa n + 2)~\vert 2n+1 \rangle $$
coincides with the Hamiltonian of a two body system
embedded in the Calogero-Sutherland of the form
$$
V_1(x,\kappa) = \frac{\kappa^2}{4} x^2 -
                 \frac{(1-\kappa)(3\kappa -1)}{4\kappa^2}
   \frac{1}{x^2} +
   \frac{1}{2}.
$$
Using the super-symmetric quantum mechanics techniques, it is simply verified that  $V_1(x,\kappa)$ is the super-symmetric partner of $
V_0(x,\kappa)$ in accordance with the fact that the operators $f^+f^-$ and $f^-f^+$ are isospectral. The first operator has the following sequence
of eigenvalues $ 0, 1, 2\kappa, 2\kappa + 1, 4\kappa, 4\kappa + 1, \cdots,   $
and for  $f^-f^+$ we have $ 1, 2\kappa, 2\kappa + 1, 4\kappa, 4\kappa + 1, \cdots,   $. The
energy levels are not equidistant and the energy gaps between two successive states are $1$ or $2\kappa - 1$.

In other hand, the algebra ${\cal B}_{\kappa}(1)$ can be related to the generalized Weyl-Heisenberg ${\cal A}_{\kappa} (1)$ which
has been shown useful in describing some exactly solvable Hamiltonians \cite{Daoud2,Daoud3}.  For
this purpose, we define the operators
$$X^+ = (f^+)^2 \qquad  X^- = (f^-)^2 $$
in terms of the creation and annihilation operators $f^+$ and $f^-$. Clearly, this realization is not possible
with usual creation and annihilation fermionic operators $(\kappa = 0)$  satisfying the relations $(f^+)^2 = 0$ and $(f^-)^2 = 0$. In view
of the equation (\ref{action-}) and (\ref{action+}), we have
$$ X^+  \vert n \rangle = \sqrt{F_+ (n+2) F_+ (n+1)} \vert n+2\rangle   \qquad  X^-  \vert n \rangle = \sqrt{F_+ (n) F_+ (n-1)}\vert n-2\rangle. $$
Then, one can write the operator products  $X^+X^-$ and $X^-X^+$  as follows
$$X^+X^- = F_+ (N) F_+ (N-1) \equiv F(N)  \qquad X^-X^+ = F_+ (N+2) F_+ (N+1) \equiv F(N+2). $$
where the structure function $F_+ (N)$ is given by (\ref{F_+(N)}).  The expression of the new structure function $F(N)$ reads
$$F(N) = \kappa^2 N(N-1) + \kappa(\kappa -1) N -  \kappa(\kappa -1) \frac{1 - (-)^{N}}{2},$$
and one has the following commutation relation
$$[ X^- , X^+ ] = 2\kappa (2\kappa N + 1)$$
to be compared with one satisfied by the ladder operators of the generalized Weyl-Heisenberg algebra. In fact, setting
$$ a^{\pm} = \frac{X^{\pm}}{\sqrt{2\kappa}},$$
one finds
\begin{equation} \label{akappa}
[ a^- , a^+ ] = 2\kappa N + 1  \qquad [N , a^{\pm} ] = \pm 2 a^{\pm}.
\end{equation}
corresponding to the structure relations defining the algebra $A_{\kappa} (1)$. This mapping  to pass from the
generalized fermion algebra to generalized boson algebra provides another example of the relevance
of the algebra ${\cal B}_{\kappa}(1)$ in the context of quantum mechanical systems possessing quadratic spectrum.
In fact, the connection between the algebra defined by (\ref{akappa}) and one-dimensional solvable potentials such as
Morse or P\"oschl-Teller potentials was discussed in \cite{Daoud3}. Accordingly, the generalized fermion algebra ${\cal B}_{\kappa}(1)$
provides the algebraic framework to deal with exactly solvable potentials with linear as well as quadratic spectra.



\section{ Coherent states and Bargmann representation}

For the generalized fermion algebra ${\cal B}_{\kappa}(1)$ a realization \`a la Bargmann can be found. This realization is the
extended version of Bargmann representation for the usual fermions. In general, the derivation of these representations is the same
as the usual fermion but the notion of the derivative and consequently the integral should be redefined. The appropriate  way to approach
the Bargmann representation for the algebra  ${\cal B}_{\kappa}(1)$ is the coherent states formalism. A coherent state for
a generalized fermion is defined as the  eigenvector of the annihilation operator
\begin{equation} \label{cs}
f^- |z \rangle  = z |z \rangle
\end{equation}
where $z$ is a complex variable. We shall assume for now that $\kappa \neq 0$. The special  case $ \kappa = 0$ where the Fock space has two dimensional basis will be discussed
hereafter as a limiting case. To anticipate,  we shall show that the bosonic variable $z$ tends to a Grassmann variable to obtain the usual coherent states
for an ordinary fermion.  To solve the  the eigenvalue equation (\ref{cs}), we
expand the state $|z \rangle$ in the Fock space basis $\vert n \rangle$
\begin{equation} |z \rangle = \sum_{n=0}^{\infty} c_{2n} ( z) |2n
\rangle +  \sum_{n=0}^{\infty} c_{2n+1 } ( z) |2n+1 \rangle.
\end{equation}
Substituting this expression in (\ref{cs}), we get the following recurrence relation
\begin{equation}
 z ~ c_{2n}( z)  = \sqrt{ 1 + 2 \kappa n } ~ c_{2n+1} (z) \qquad  z ~ c_{2n+1} = \sqrt{ 2 \kappa (n+1) }~  c_{2n+2} (z)
( z) \end{equation}
which leads to
\begin{equation} \label{coef.c}
c_{2n} ( z) = c_0 ( z ) \frac{ z^{2n} }{ \sqrt{ (2\kappa )^{2n} n!  \Gamma (\frac{1}{2 \kappa} +  n).
}}\qquad c_{2n+1} ( z) = c_0 ( z ) \frac{ z^{2n+1} }{ \sqrt{ (2\kappa )^{2n+1} n!  \Gamma (\frac{1}{2 \kappa} +  n +1).
}}.
\end{equation}
As a result, we get the normalized  coherent state vectors
\begin{equation} \label{cs-eq}
|z \rangle = \bigg[ {\cal N}_{\kappa} (|z|^2) \bigg]^{-\frac{1}{2}} \sum_{n= 0}^{\infty}
\frac{ z^{2n} }{ \sqrt{ (2\kappa )^{2n} n!  \Gamma (\frac{1}{2 \kappa} +  n)}}\bigg[ | 2n \rangle + \frac{z}{\sqrt{1+2\kappa n}} | 2n+1 \rangle\bigg].
\end{equation}
The normalization factor ${\cal N}_{\kappa} (|\psi|^2)$ is given by
$$
 {\cal N}_{\kappa} (
|z|^2 ) =  e_{\kappa} \left( \frac{|z|^2 }{2
\kappa }\right)
$$
where the function $e_{\kappa}(.)$ is defined by
$$
e_{\kappa} (x) = \sum_{n=0}^{\infty} \bigg(\frac{1}{2 \kappa} +  n + x \bigg) \frac{x^{2n}}{n! \Gamma (\frac{1}{2 \kappa} +  n +1)}.
$$
The coherent states provide a polynomial realization of the algebra ${\cal B}_{\kappa}(1)$. Indeed, Any state $\vert \Psi \rangle$
$$ \vert \Psi \rangle = \sum_{n=0}^{\infty}  \bigg( \Psi_{2n}\vert 2n \rangle + \Psi_{2n+1}\vert 2n+1 \rangle \bigg) $$
is represented, in the Bargmann representation, as a function of the complex variable $z$ according the following correspondence
\begin{equation} \label{psi(z)}
\vert \Psi \rangle \longrightarrow \Psi(z) = \bigg[ {\cal N}_{\kappa} (|z|^2) \bigg]^{\frac{1}{2}} \langle \bar z \vert \Psi\rangle = \sum_{n=0}^{\infty}  \bigg( \Psi_{2n} f_{2n}+ \Psi_{2n+1} f_{2n+1}\bigg)
\end{equation}
where the monomials $f_{2n}(z)$ and $f_{2n+1}(z)$ are
\begin{equation} \label{f2n}
 f_{2n} (z) = \bigg[ {\cal N}_{\kappa} (|z|^2) \bigg]^{\frac{1}{2}}\langle \bar z \vert 2n \rangle =   \frac{ z ^{2n} }{\sqrt{ (2\kappa )^{2n} n! \Gamma (\frac{1}{2 \kappa} +  n ) }}
\end{equation}
and
\begin{equation} \label{f2n+1}
 f_{2n+1} (z) =  \bigg[ {\cal N}_{\kappa} (|z|^2) \bigg]^{\frac{1}{2}}\langle \bar z \vert 2n+1 \rangle =   \frac{ z ^{2n+1} }{\sqrt{ (2\kappa )^{2n+1} n! \Gamma (\frac{1}{2 \kappa} +  n +1)}}
\end{equation}
correspond  to the analytical representation of the vectors $\vert 2n \rangle $ and $\vert 2n+1 \rangle$ respectively. In this respect, the Fock space
spanned by the number basis $\{ \vert n \rangle, n = 0, 1, 2, \cdots\}$ is equivalent to the space of
 functions of the variable $z$ spanned by the basis $\{ f_{0} (z) , f_{1} (z) , f_{2} (z), \cdots\}$. Furthermore, using the equations (\ref{cs-eq}), (\ref{f2n}) and (\ref{f2n+1}), we realize the
creation, annihilation and number operators, in the Bargmann representation, as follows
$$ f^+ \equiv z \qquad f^- \equiv \frac{1}{z} F_+\bigg(z\frac{d}{dz}\bigg) \qquad N \equiv z\frac{d}{dz}$$
where $F$ is the structure function defined by (\ref{F_+(N)}) or equivalently by (\ref{F_+(N)1}). Using the expressions (\ref{f2n}) and (\ref{f2n+1}), it is simple to check that
$$ N ~ f_{2n} (z) =  2n ~ f_{2n}(z) \qquad N ~ f_{2n+1} (z) = (2n+1)~  f_{2n+1}(z). $$
The creation operator acts as multiplication by $z$ and gives
$$ f^+ ~ f_{2n} (z) = \sqrt{1 + 2\kappa n} ~ f_{2n+1}(z) \qquad f^+ ~ f_{2n+1} (z) = \sqrt{2\kappa (n+1)}~  f_{2n+2}(z). $$
To write the explicit from of the action of the annihilation operator, we note that the action of the operator $(-)^N$
on any analytical function of the form (\ref{psi(z)}) induces the parity transformation $z \longrightarrow -z $.
Then, using the result (\ref{F_+(N)1}),  the action of the annihilation operator $f^-$ can be represented as
\begin{equation} \label{Dkappa}
D^{\kappa}_{z} =  \kappa \frac{d}{dz} + (1-\kappa) \frac{\partial}{\partial_{-1}z}
\end{equation}
where the second  object in the right-hand side is defined by
\begin{equation} \label{finiteD}
 \frac{\partial}{\partial_{-1}z} = \frac{f(z) - f( - z)}{ 2z}
\end{equation}
 is a special form  of the Fibonacci difference operator \cite{Arik2}. Using (\ref{Dkappa}) together with (\ref{finiteD}), one obtains
$$ f^- ~ f_{2n} (z) = \sqrt{2\kappa n} ~ f_{2n-1}(z) \qquad f^- ~ f_{2n+1} (z) = \sqrt{1 + 2\kappa n}~  f_{2n}(z). $$
reflecting that the differential operator  $ D^{\kappa}_{z} $
acts as a generalized derivative in the Bargmann representation of the generalized fermion algebra ${\cal B}_{\kappa}(1)$. It is easy
to check the anticommuation relation
\begin{equation} \label{anti}
\{ \frac{\partial}{\partial_{-1} z }, z \} =1
\end{equation}
which is reminiscent of Grassmann variables. The generalized derivative $ D^{\kappa}_{z}$ is the sum of two derivatives: the usual(bosonic) one and the second is
of Grassmann type. Finally, using the relation (\ref{anti}), one has
\begin{equation}
\{ D^{\kappa}_{z} , z \} = 1 + 2 \kappa z \frac{d}{d z }.
\end{equation}
Now, we consider the special case $\kappa = 0$. It is important to emphasize that for $\kappa \to 0$, the coefficients $c_{2n}(z)$ and $c_{2n+1}(z)$,  given by (\ref{coef.c}), go to infinity for $n > 0$. Therefore, the terms in $z^{2n}$ $(n>0)$ in Eq. (\ref{cs}) make
sense  only if $z$ is a  Grassmann variable with
$$ z^2 = 0.$$
Hence, the coherent state (\ref{cs}) reduces to
$$ \vert z \rangle = \frac{1}{ (1 + z\bar z)^{\frac{1}{2}}} [ \vert 0 \rangle + z \vert 1 \rangle ],$$
and the Fock basis corresponds
$$ \vert 0 \rangle \longrightarrow  f_0(z) = 1 \qquad  \vert 1 \rangle \longrightarrow  f_1(z) = z . $$
In this limiting case, the generalized derivative  $D^{\kappa}_{z}$ (\ref{Dkappa}) coincides with the partial derivative of the Grassmann variable $z$ and satisfies the nilpotency
condition
$$ \bigg(\frac{\partial}{\partial_{-1} z }\bigg)^2 = 0$$
because the Bragmann space has a two dimensional basis $\{ 1 , z \}$.


\section{ Normal ordering process}

In quantum field theory a product of quantum fields, or equivalently
their creation and annihilation operators, is usually said to be
normal ordered (also called Wick order) when all creation operators
are to the left of all annihilation operators in the product. The
process of putting a product into normal order is called normal
ordering (also called Wick ordering). The process of normal ordering
is particularly important for a quantum mechanical Hamiltonian. When
quantizing a classical Hamiltonian there is some freedom when
choosing the operator order, and these choices lead to differences
in the ground state energy. In this subsection we deal with the
normal ordering process for the generalized fermion algebra ${\cal B}_{\kappa}(1)$. The normal ordering is obtained by using the Wick's theorem;
\begin{equation}
( f^+ f^- )^r = \sum_{k=1}^r ( f^+)^k  S_{\kappa} (r,
k, N) ( f^-)^k
\end{equation}
where $S_{\kappa} (r,k, N)$ is the
$\kappa$-deformed Stirling operator of the second kind. Using the identity\\ $ (f^+ f^- )^{n+1} = (f^+ f^- ) (f^+ f^-)^n $, we obtain
the recurrence relation
\begin{eqnarray}\label{Srel}
&& S_{\kappa} (r+1,k, N) =
(-1)^{k-1} S_{\kappa} (r,k-1, N+1)\nonumber\\
&&  +
(-1)^{k-1} \left( \frac{ 1- (-1)^k }{2} ( 1 + 2 \kappa N ) + \kappa
k (-1)^k  + \frac{ 1- (-1)^k }{2} \kappa \right) S_{\kappa}(r,k,c)
\end{eqnarray}
The first few  $\kappa$-deformed   Stirling operator of the second
kind are \begin{equation}
 S_{\kappa}(1,1, N) =I,
\end{equation}
 \begin{equation}
 S_{\kappa}(2,1,N) =  I + 2 \kappa N  , ~~~ S_{\kappa}(2,2,N) =-I
\end{equation}
 \begin{equation}
 S_{\kappa}(3,1,N) =( I + 2 \kappa N )^2  , ~~~ S_{\kappa}(3,2,N) =-I - 2 \kappa N , ~~~ S_{\kappa}(3,3,N) = I
\end{equation}
 \begin{equation}
   S_{\kappa}(4,1,N) = (I + 2 \kappa N) ^3 , ~~~ S_{\kappa}(4,2,N) =- ( I -2 \kappa + 4 \kappa^2 ) - 4 \kappa ( 1 + \kappa ) N - 4 \kappa^2 N^2 ,
     \end{equation}
   \begin{equation}
    ~~~ S_{\kappa}(4,3,N) = - 4 \kappa  , ~~~ S_{\kappa}(4, 4,N) = -I.
 \end{equation}
 The $\kappa$-deformed Bell operator is then defined  as
\begin{equation} \label{B}
B_r(N) = \sum_{k=1}^r S_{\kappa} ( r, k, N ).
\end{equation}
Some of the $\kappa$-deformed Bell operators are:
\begin{equation}
B_1 (N) = I
\end{equation}
\begin{equation} B_2 (N) = 2\kappa N
\end{equation}
\begin{equation}
B_3 (N) = I + 2 \kappa N + 4 \kappa^2 N^2
\end{equation}
\begin{equation}
B_4 (N) = - ( I+2 \kappa + 4 \kappa^2 ) +2  \kappa ( 1-2 \kappa ) N +8
\kappa^2 N^2 + 8\kappa^3 N^3 ,
\end{equation}
Finally, in the limit $\kappa \rightarrow 0 $, using the equations (\ref{Srel}) and (\ref{B}), one gets
\begin{equation}
B_1(N) = I , ~~~ B_2 (N) =0,
\end{equation}
and for $ r \geq 3$, one shows
\begin{equation}
B_r (N) =
\cases{ (-1)^r I ~~( r=0 ~ (\mbox{mod} ~ 3) ) \cr
 (-1)^{r+1} I ~~( r=1 ~ (\mbox{mod} ~ 3) ) \cr
 0 ~~( r=2 ~ (\mbox{mod} ~ 3) ) }.
\end{equation}
which correspond to the Bell operators for ordinary fermion algebra.

\section{Concluding remarks and perspectives}

The main idea of this work is the "{\it fermionization}" of the generalized Weyl-Heisenberg algebra ${\cal A}_{\kappa}(1)$
investigated in \cite{Daoud3}. This "{\it fermionization}" procedure is  introduced by simply replacing the commutator of
 creation and annihilation operators, in the structure relations of ${\cal A}_{\kappa}(1)$, by the anti-commutator. The
resulting algebra ${\cal B}_{\kappa}$ covers the ordinary fermion algebra.  In contrast with the several extensions of fermionic algebra
introduced in the context of quantum algebras \cite{Bona}, the algebra ${\cal B}_{\kappa} (1)$ is not equivalent to the usual fermionic algebra. The
parameter $\kappa$  induces a drastic deviation from  fermions  to  bosons. Indeed, the Fock and Bargmann realizations show clearly that
the representation spaces is infinite dimensional and become suddenly two dimensional when $\kappa$ approaches zero. This deviation also arises
in the context of the coherent states constructed in this work. Indeed, we have shown that in the limiting situation $\kappa \rightarrow 0$, they reduce
to fermionic coherent states involving Grassmann variables. In view of the connection between the algebra ${\cal B}_{\kappa}(1)$ and the one-dimensional Calogero-Sutherland system
presented in this work, it is worthwhile to look for the appropriate scheme to generalize ${\cal B}_{\kappa} (1)$  to describe other potentials. Also, we notice that
 the representation of ${\cal A}_{\kappa}(2)$ algebra \cite{Daoud4} was generalized
to incorporate multi-bosons \cite{Chung2}.  Hilbertian as well as analytical representations were discussed.  In this respect, it is
 interesting to investigate the multimode generalization of the generalized fermion algebra ${\cal B}_{\kappa}(1)$. Some preliminaries results,
regarding this issue,  were obtained \cite{CD}
and we hope report on this subject in a forthcoming paper.

\end{document}